\pdfoutput=1
\documentclass[aps,prc,twocolumn,superscriptaddress]{revtex4-2}

\usepackage{graphicx}  
\usepackage{amsmath}
\usepackage{amsfonts}
\usepackage{amsbsy}
\usepackage{bm}  %
\usepackage{color}
\usepackage{xcolor}
\usepackage{hyperref}
\hypersetup{
    colorlinks,
    linkcolor={red!50!black},
    citecolor={blue!50!black},
    urlcolor={blue!80!black}
}

\def\mqo2{{\!\!\!}}

\newcommand{\cL}{\mathcal{L}}

\begin{document}
% * Title
\title{Non-Efimovian two-neutron halos with an $s$-wave core-neutron resonance}
\author{Lucas Platter}\email{lplatter@utk.edu}
\affiliation{Department of Physics and Astronomy, University of Tennessee, Knoxville, Tennessee 37996, USA}
\affiliation{Physics Division, Oak Ridge National Laboratory, Oak Ridge, Tennessee 37831, USA}

\author{Dam Thanh Son}
\affiliation{Leinweber Institute for Theoretical Physics, University of Chicago, Chicago, Illinois 60637, USA}
\date{July 2025}

\begin{abstract}
We consider two-neutron halo nuclei in which the neutron core subsystem displays a resonance close to threshold. Such resonances can be generated in an effective field theory in which the scattering length and effective range are summed to all orders. We show that no three-body parameter is required to make predictions in this case and map out the universal features of such systems. We furthermore study the dependence of these universal features on the core mass. We apply our framework to the two-neutron halo nucleus ${}^{22}$C.
\end{abstract}

\smallskip
\maketitle
%\tableofcontents
\emph{Introduction.---}Neutron halo nuclei are systems composed of a tightly bound core and one or more weakly bound neutrons. These nuclei are characterized by small one- or two-neutron separation energies and  extended matter radii. One-neutron halo nuclei consist of a neutron loosely bound to a compact core, while two-neutron halo nuclei, such as $^6$He, $^{11}$Li, $^{19}$B, and $^{22}$C, exhibit small two-neutron separation energies and are typically understood as shallow three-body bound states of a core and two neutrons~\cite{Tanihata_1996}.  These nuclei are examples of Borromean systems, in which no two-body subsystem is bound, yet the three-body system is.

A remarkable feature of some two-neutron halo nuclei, notably $^{11}$Li and $^{19}$B, is a large negative $s$-wave scattering length in the core-neutron ($cn$) subsystem.  Such systems have attracted significant theoretical interest because they can, in certain limits, be thought of as Efimov states~\cite{Braaten:2004rn} that arise when the pairwise $s$-wave scattering lengths are large compared to the interaction range. Such systems exhibit a discrete scaling symmetry in the limit of vanishing interaction range. A crucial feature of their theoretical description is the need for a single three-body datum to render predictions finite and regulator-independent. In Efimov’s original hyperspherical formalism, this datum is a boundary condition at short distances~\cite{PhysRevC.44.2303}. In the effective field theory (EFT) framework developed by Bedaque, Hammer, and van Kolck~\cite{Bedaque:1998kg}, the three-body datum enters as a coupling constant for a leading-order three-body force.

This EFT approach was originally applied to systems of three identical atoms~\cite{Bedaque:1998kg} and the three-nucleon system~\cite{Bedaque:1999ve}. However, in recent years, it has also been extensively applied to halo nuclei~\cite{Hammer:2017tjm} and is called Halo EFT in these applications.

It is not obvious that all two-neutron halos have to have Efimovian character.  In the $^6$He nucleus there is no $s$-wave resonance in the alpha-neutron subsystem, but a $p$-wave resonance.  In the $^{22}$C nucleus, one experiment indicates that the core-neutron scattering length is rather small~\cite{Mosby:2013bix}. 
%while other data~\cite{Leblond:2015} indicate a $s$-wave resonance at energy of order $1$~MeV with a similar width.
Motivated by these situations, Hongo and Son constructed an EFT for weakly bound halo nuclei with no $s$-wave $cn$ resonance~\cite{Hongo:2022sdr}. 
%This theory has two fine-tuning parameters: the large neutron-neutron $S$-wave scattering length and the small two-neutron separating energy of the halo nucleus.  
This model has a logrithmically running coupling and needs a cutoff at, or below, the scale of the Landau pole of this coupling.

In this paper, we explore non-Efimovian halo nuclei that appear in a simple field-theoretical model that is well-defined to arbitrarily small length scales and with only two-body input.  In that context, we recall a paper by Petrov~\cite{PhysRevLett.93.143201} which shows that, in the case of three identical bosons, no three-body parameter is needed when the two-body interaction supports a near-threshold resonance with a large negative effective range in addition to a large scattering length. Subsequent work developed EFTs for the two‑body~\cite{Habashi:2020ofb,Habashi:2020qgw} and three‑body~\cite{Nishida:2012xw,Griesshammer:2023scn} sectors, for systems of identical particles.

This scenario is particularly relevant in atomic systems near a narrow Feshbach resonance and opens the possibility of a new type of three-body universality distinct from the Efimov scenario. However, while this regime has been explored in atomic systems near narrow Feshbach resonances, its implications in nuclear few-body systems remain underexplored.

The core-neutron-neutron ($cnn$) problem involves two distinct two‑body channels: the $cn$ system, which may possess a resonance, and the neutron-neutron ($nn$) system, which does not. It is unclear a~priori whether a large negative $cn$ effective range alone suffices to renormalize the three‑body equations without an additional parameter.  Here, we will demonstrate that, under such conditions, the $cnn$ system remains well-defined and does not require a three-body force for renormalization. Furthermore, we will map out the region in the space of $cn$ scattering lengths and effective ranges where a three-body bound state appears.

%This paper is organized as follows. We start by discussing the S-matrix structure for negative scattering length and effective range. We then introduce the Lagrangian and derive the integral equations used in our analysis. We present and discuss our numerical results. Finally, we summarize our findings and outline possible future directions.

\emph{The core-neutron system.}---The effective range expansion is the Taylor expansion
\begin{equation}
k \cot \delta = -\frac{1}{a} + \frac{r}{2} k^2 +\ldots~,
\end{equation}
in even powers of the relative momentum $k$ between two scattered particles, where $\delta$ denotes the scattering phase shift. The first two parameters in this expansion are the scattering length $a$ and effective range $r$.

We consider the scenario in which both the $cn$ scattering length $a_{cn}$ and effective range $r_{cn}$ are negative and higher coefficients can be neglected. In this scenario, two cases can be distinguished: (i) $2 r_{cn} < a_{cn}$ and (ii) $2 r_{cn} > a_{cn}$. In case (i), we have two resonance poles at $k_{\pm} = \pm k_R - i k_I$, where $k_R = -\sqrt{2 r_{cn}/a_{cn} -1 }/r_{cn}$ and $k_I = -1/r_{cn}$. The impact of this pole structure can be seen, for example, in the elastic $cn$ scattering cross section.  In case (ii), we have two virtual states that are reflected by two poles on the negative imaginary axis with
$k_{\pm} = i (1\pm \sqrt{1-2 r_{cn}/a_{cn}})/r_{cn}$. In this case, we find an enhancement of the cross section at threshold.

\begin{figure}[t]
  \centerline{\includegraphics[width=0.9\columnwidth]{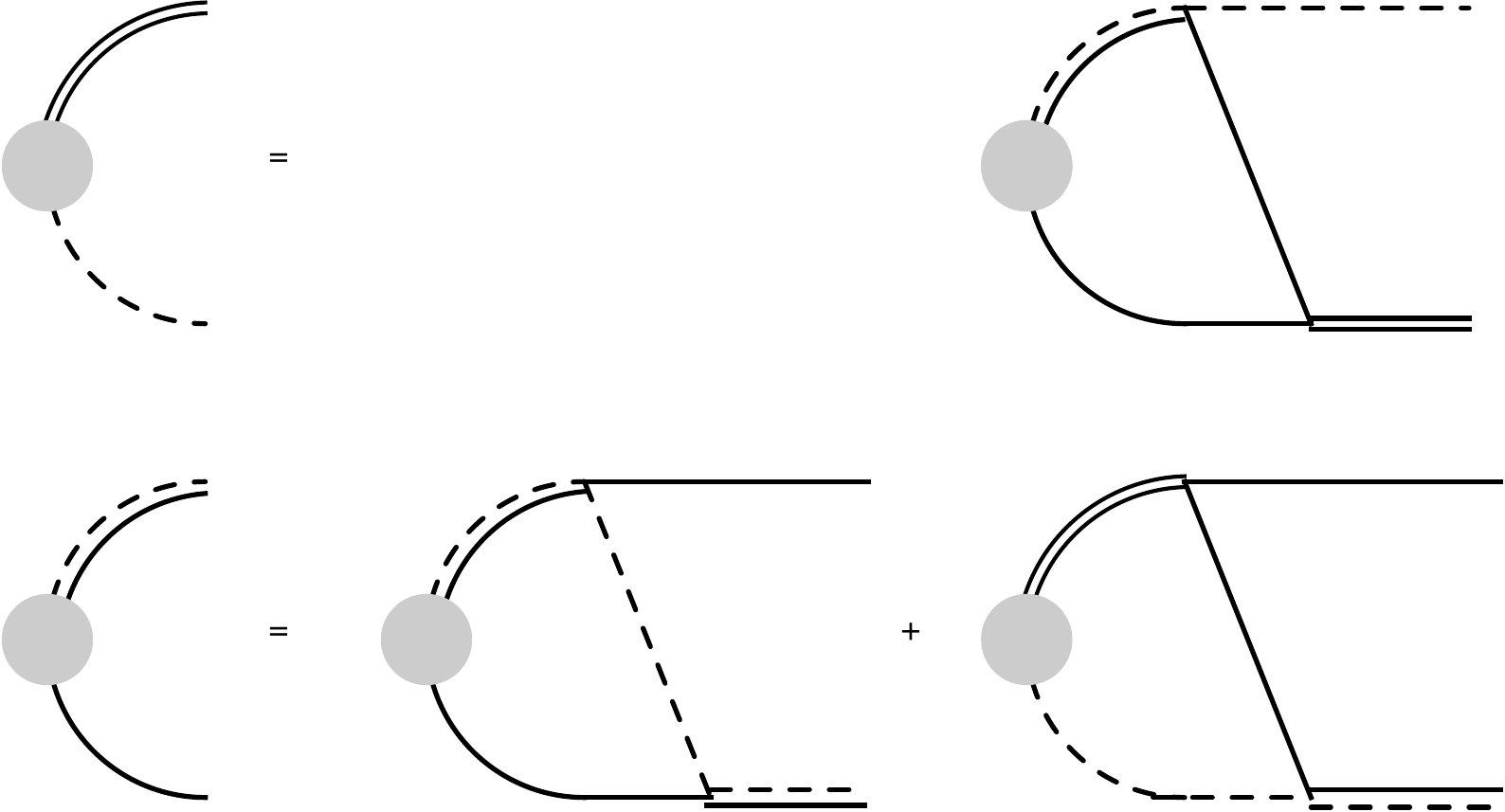}}
  \caption{Diagrammatic form of the integral equation given in Eq.~\eqref{eq:inteq}. The gray circle denotes the vertex function ${\bf G}$, the solid line a neutron propagator, and the dashed line a core propagator. The double lines represent the corresponding dimer propagators.}
  \label{fig:halo-feynman}
\end{figure}

\emph{Two-neutron systems with a large effective range.}---%
%\label{sec:inteq}
In order to calculate properties of the three-body system with a large $cn$ effective range, we use a modified version of the Lagrangian used in Halo EFT~\cite{Vanasse:2016hgn}
\begin{multline}
%\nonumber
    \cL = c^\dagger\biggl(i \partial_0 +\frac{\nabla^2}{2 A m}\biggr)c +n^\dagger\biggl(i\partial_0 +\frac{\nabla^2}{2m}\biggr) n\\
+ \Delta_0 d_0^\dagger  d_0
-\sqrt{\frac{\pi}{2m}}\bigl(d_0^\dagger n^Ti \sigma_2 n +\rm{h.c.}\bigr) 
\\
%\nonumber
+ d_1^\dagger \biggl[\Delta_1 + i \partial_0 +\frac{\nabla^2}{4(A+1)m}
\biggr] d_1
%\\
%&
- g_1 \bigl(d_1^\dagger n c + \rm{h.c.}\bigr),
\end{multline}
where $c$ denotes the core field, $n$ the neutron field, $d_0$ and $d_1$ denote $nn$ and $cn$ dimer fields, $m$ is the  nucleon mass, and $A$ is the ratio of the core mass $m_c$ over the nucleon mass, i.e., $A=m_c/m$. The parameter $\Delta_0$ is renormalized such that the $nn$ scattering amplitude has a virtual state resulting from a large negative scattering length $a_{nn}$.  We renormalize the parameters  $\Delta_1$ and $g_1$ to reproduce the $cn$ scattering length $a_{cn}$ and effective range  $r_{cn}$. 

We obtain the dressed dimer propagators by summing up the neutron-neutron loops and core-neutron loops to all orders. The renormalized two-neutron dimer propagator with energy $q_0$ and momentum ${\bf q}$ us
\begin{equation}
D_{nn}(q_0, {\bf q}) = \biggl(\sqrt{-m q_0 +\frac{{\bf q}^2}{4}-i\epsilon}-\frac{1}{a_{nn}}\biggr)^{-1} ,
\end{equation}
%where $a_{nn}$ denotes the neutron-neutron scattering length. 
while the propagator of the $cn$ dimer is
\begin{multline}
    D_{cn}(E,q) = \Biggl[\sqrt{-\frac{2A}{1+A} m q_0+\frac{A}{(A+1)^2}{\bf q}^2-i\epsilon}\\
    -\frac{1}{a_{cn}}+\frac{r_{cn}}{2}\left(\frac{2 A}{1+A}m q_0-\frac{A}{(1+A)^2}q^2\right)\Biggr]^{-1}~.
\end{multline}
\begin{figure}[t]
  \centerline{\includegraphics[width=0.9\columnwidth]{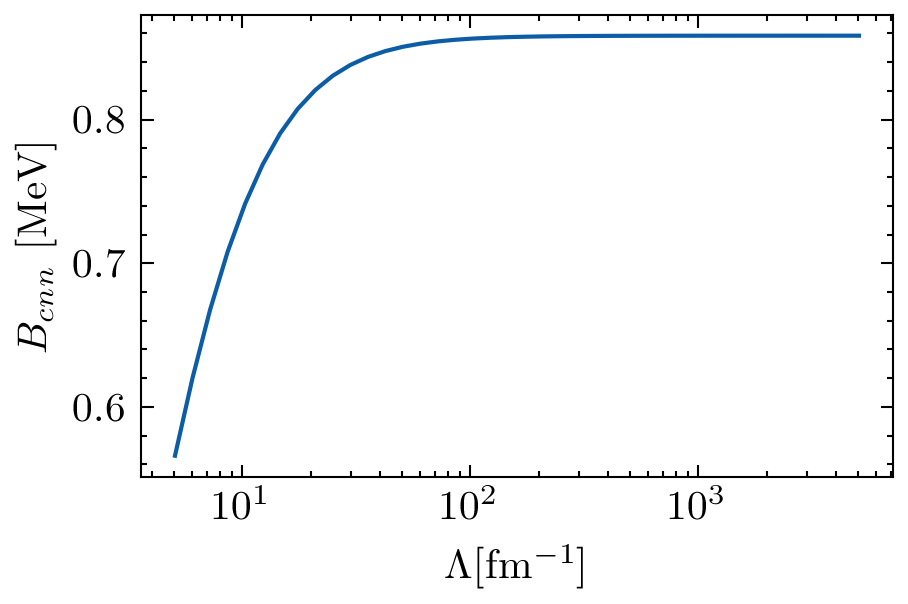}}
  \caption{\label{fig:cut_dep_unitarity} Cutoff dependence of the binding energy of a three-body state with an infinite $nn$ scattering length,
%  $a_{cn} = 0.5$~MeV$^{-1}$ and $r_{cn} = 0.3$~MeV$^{-1}$.
  $a_{cn} = -5$~fm, and $r_{cn} = -3$~fm.  Here $A=20$.}
\end{figure}
The $cnn$  system is described by the integral equation shown diagrammatically in Fig.~\ref{fig:halo-feynman}. It is a coupled integral equation whose solution gives two amplitudes that are part of the quantity ${\bf G}$
\begin{equation}
\label{eq:inteq}
    {\bf G}(E, p) = {\bf R}(E,p,q){\bf D}(E,q) \otimes_q G(E,q)~.
\end{equation}
where the operator $\otimes_q$ indicates an integral over the momentum $q$
\begin{equation}
    A(q)\otimes_q B(q) = \frac{1}{2\pi^2}\int_0^\Lambda dq q^2 A(q)B(q)~,
\end{equation}
and the matrix function ${\bf R}(E,p,q)$ that gives the $s$-wave projected single particle exchange is given by
\begin{align}
    {\bf R}(E,p,q) = \begin{pmatrix}
    R_{00}(E,p,q) & R_{01}(E,p,q)\\
    R_{10}(E,p,q) & R_{11}(E,p,q)
    \end{pmatrix} .
\end{align}
For the elements of this matrix function, we have
\begin{equation}
    R_{01}(E,p,q)= R_{10}(E,q,p)~,
\end{equation}
\begin{align}
    R_{01}(E,p,q) = \frac{2\sqrt{2}\pi(1{+}A)}{A\, p q}
    Q_0\biggl(\frac{\frac{1+A}{2A} p^2 {+} q^2 {-} m E}{p q}\biggr),
\end{align}
and
\begin{align}
    R_{11}(E,p,q) &= \frac{(1{+}A)^2}{A}\frac{\pi}{p q}
    Q_0\biggl( \frac{\frac{1+A}{2}(p^2{+}q^2){-} A m E}{p q}\biggr),
\end{align}
where $Q_0(x)$ is the Legendre function of the second kind.
The integral equation \eqref{eq:inteq} is solved by first discretizing the momentum and then solving for the roots of the characteristic polynomial of the resulting matrix. 

\begin{figure}[t]
  \centerline{\includegraphics[width=0.9\columnwidth]{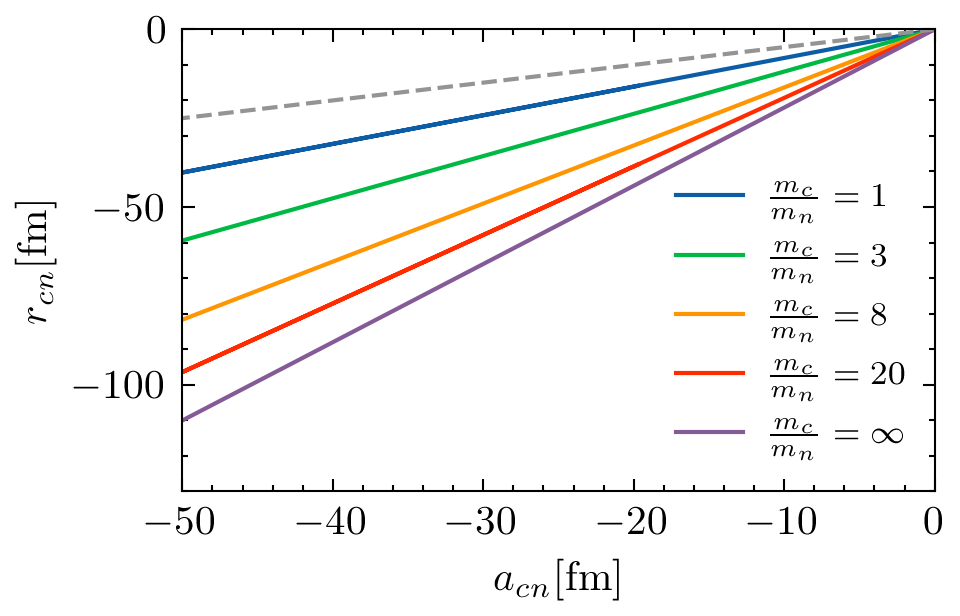}}
  \caption{The combination of scatterings length $a_{cn}$ and $r_{cn}$ that lead to a three-body state at threshold.}
  \label{fig:b3_threshold_unitarity}
\end{figure}
\emph{Results.}---%
%\label{sec:results}
%We have solved the integral equation in Eq.~\eqref{eq:inteq} numerically. 
We consider two different cases for the neutron-neutron interaction: (i) infinite scattering length, when the $nn$ system is at unitarity, and (ii) physical scattering length $a_{nn}$, chosen to reproduce a virtual pole in the dimer propagator at $1/
a_{nn} = -9.87~\text{MeV}/(\hbar c)$. In Fig.~\ref{fig:cut_dep_unitarity}, we show the cutoff dependence of the binding energy of a $cnn$ state. The binding energy converges rapidly with the cutoff, indicating that no additional three-body parameter is required.  We have also performed this analysis using the physical value of the $nn$ scattering length and observe the same qualitative behavior: no additional three-body parameter is required for renormalization.

\begin{figure}[t]
  \centerline{\includegraphics[width=0.9\columnwidth]{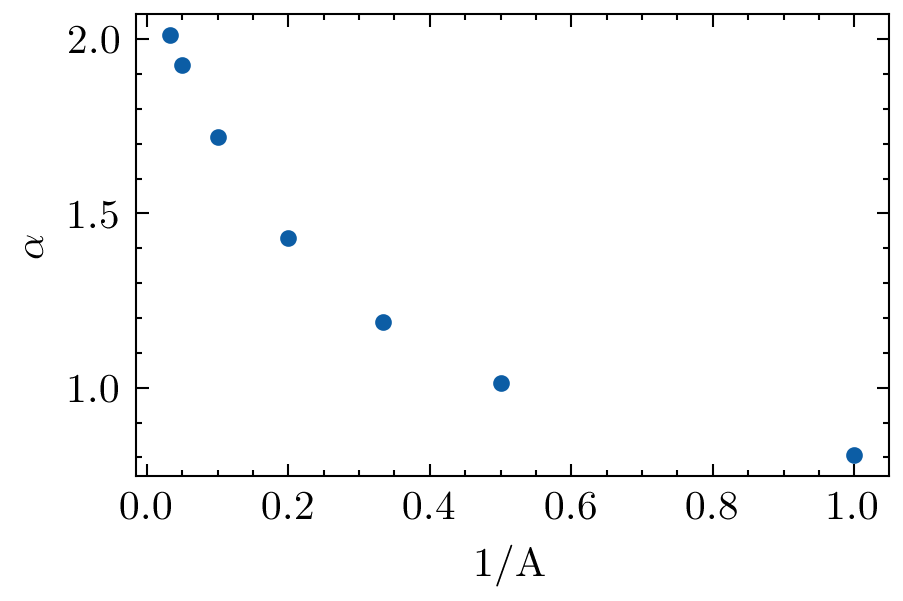}}
  \caption{The slope $\alpha$ that relates $a_{cn}$ to $r_{cn}$ in the limit of unitary $nn$ scattering length as a function of $1/A$~\cite{slope}.}
  \label{fig:a_vs_r_slope}
\end{figure}

Our approach is expected to be valid when the absolute values of the core-neutron scattering length and effective range are both larger than the range of the underlying interaction or, equivalently, larger than all subleading effective-range parameters.  When information about the core-neutron interaction is limited, one can take the presence of a $cn$ $s$-wave resonance (as opposed to a virtual state) and a three-body $cnn$ bound state as a signature of a non-Efimovian two-neutron halo nucleus. It is, however, a priori not clear that a $cnn$ bound state can exist when the $cn$ system displays a resonance. A core–neutron resonance is present when $2|r_{cn}|>|a_{cn}|$.  The same parameter $r_{cn}$, however, also acts as the ultraviolet regulator of the three‑body integral equations: the larger $|r_{cn}|$ is, the lower the corresponding cutoff $\Lambda\!\sim\!1/|r_{cn}|$.  
If $|r_{cn}|$ becomes too large, $\Lambda$ drops below the momentum scale needed to bind two neutrons to the core, and the $cnn$ system stays unbound despite the existence of the two‑body resonance.  Thus, a three‑body state can form only in when $|r_{cn}|$ is smaller than a certain critical value.  The question is whether that critcal value is such that there is still room for the condition $2|r_{cn}|>|a_{cn}|$ to be satisfied.

Figure~\ref{fig:b3_threshold_unitarity} shows the relation between the core-neutron scattering length and effective range for which a three-body state lies exactly at threshold, assuming that $a_{nn}=\infty$. In this limit, the only relevant dimensionless parameters are the mass number of the core $A = m_c/m$ and the ratio $a_{cn}/r_{cn}$. Consequently, for a given halo nucleus, the threshold condition defines a linear relation between $a_{cn}$ and $r_{cn}$: $r_{cn} = \alpha a_{cn}$. The dark gray dashed line in Fig.~\ref{fig:b3_threshold_unitarity} separates the regions where the $cn$ subsystem exhibits a virtual state (upper right) or a resonance (lower left). We find that, for $a_{nn} = \infty$, there exist a range of $|r_{cn}|$, $\frac12|a_{cn}|<r_{cn}<\alpha|a_{cn}|$, where the $cn$ subsystem has a resonance and the $cnn$ system is bound.  In particular, when the three-body state lies at threshold, the $cn$ subsystem has a resonance. The slope $\alpha$, which determines how wide this interval is, depends on $A$, but is always larger than $\tfrac12$. We plot $\alpha$ as a function of $1/A$ in Fig.~\ref{fig:a_vs_r_slope}. 
%The numerical results for the slopes shown in this figure are given in~\cite{slope}.

\begin{figure}[t]
  \centerline{\includegraphics[width=0.9\columnwidth]{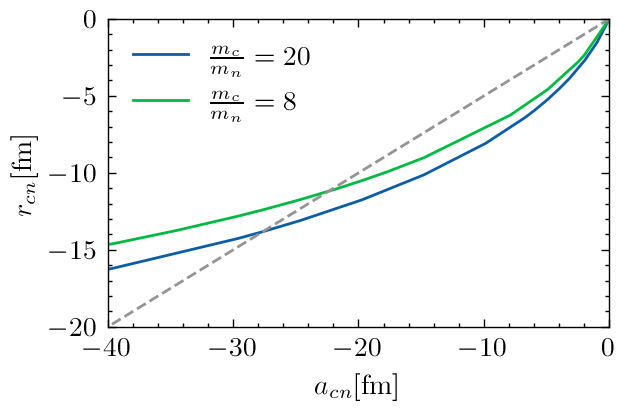}}
  \caption{The combination of scatterings length $a_{cn}$ and $r_{cn}$ that lead to a three-body state at threshold with $nn$ scattering length to its physical value. The dark grey dashed line separates the regions for which the two-body state is a virtual state (upper right) or a resonance (lower left).}
  \label{fig:b3_threshold_physical}
\end{figure}

The situation is more complicated when the $nn$ scattering length is set to its physical value. Figure~\ref{fig:b3_threshold_physical} shows the results for this case. The border between regions with and without theree-body bound state on the ($a_{cn}$, $r_{cn}$) plane is no longer a straight line due to the presence of the additional scale $a_{nn}$.  Where the three-body state is exactly at threshold, we find a $cn$ resonance for sufficiently small $cn$ scattering lengths. When $a_{cn}$ is larger than some value dependent on $A$ (and of order $a_{nn}$), the two-body system possesses a virtual state instead.
In both cases, the three-body state at threshold is the only bound state in the system.

\begin{figure}[t]
  \centerline{\includegraphics[width=0.9\columnwidth]{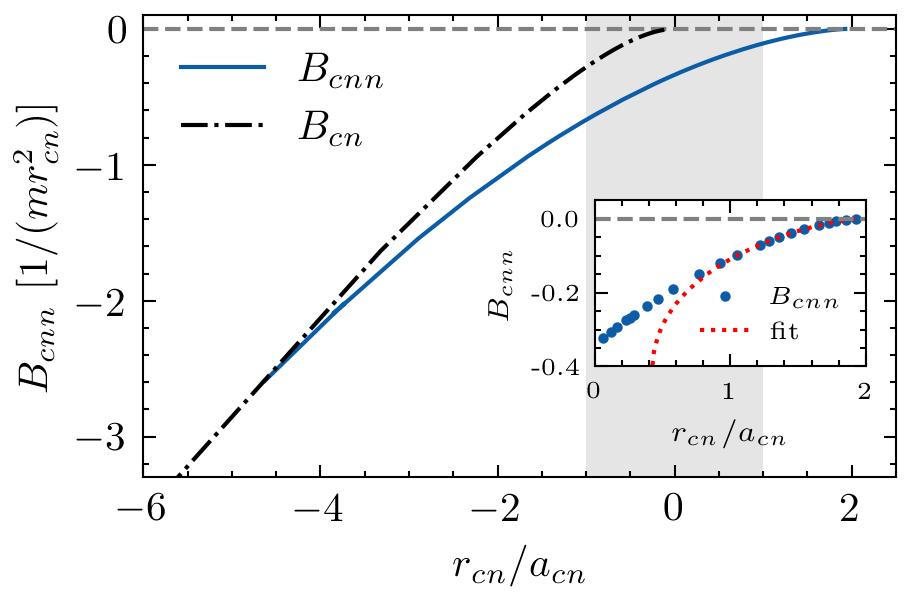}}
  \caption{The solid line gives the $cnn$ binding energy as a function of $r_{cn}/a_{cn}$ for infinite $nn$ scattering length and $A=20$. The dashed-dotted line shows $cn$ two-body binding energy. The grey shaded area gives the region that is well within the domain of standard Halo EFT. The inset shows is a magnification of threshold region. The blue dots denote the numerical data while that red dotted line shows a fit to the numerical data (see text for details).}
  \label{fig:efimov-mod} 
\end{figure}

We can also analyze within our framework how $cnn$ states approach the thresholds as $r_{cn}/a_{cn}$. This is reminiscent of the well-known Efimov plot in which the binding momentum is shown as a function of the inverse scattering length while the short-distance three-body parameter remains fixed. In Fig.~\ref{fig:efimov-mod}, we show the binding energy as a function of $r_{cn}/a_{cn}$ for infinite $nn$ scattering length and $A=20$.  The approach of $B_{cnn}$ to zero is consistent with the behavior predicted in Ref.~\cite{Konishi:2017lbg}, that is, $|B_{cnn}|\ln \frac{B_0}{|B_{cnn}|}= C(x_0-r_{cn}/a_{cn})$.  For $A=20$, a simple fit in the region $r_{cn}/a_{cn} = 1.93$ to $r_{cn}/a_{cn}\approx 1.5$ yields $x_0=1.929$, $B_0=0.4439/mr_{cn}^2$, $C=0.2093/mr_{cn}^2$. We observe that the parameters do depend on the fit interval.   This issue should be addressed with a more sophisticated fit protocol.

\emph{Application to} $^{22}$C.---The nucleus $^{22}$C binds two neutrons and a $^{20}$C core. 
%Data on the two-neutron separation energy of $^{22}$C is scarce and imprecise: $S_{2n}=0.033(325)$ MeV \cite{Wang:2021xhn}, $-0.14(46)$ MeV \cite{Gaudefroy:2012qe}.  The value of the rms matter radius is also controversial: $r_\text{rms}=5.4(9)$ fm \cite{Tanaka:2010zza} (from which a very small value $S_{2n}$ is deduced, $S_{2n}\sim0.01$ MeV), and 3.44(8) fm \cite{Togano:2016wyx}, indicating a much larger $S_{nn}\sim0.4$ MeV.  
%The situation with the core-neutron subsytem: 
While Ref.~\cite{Mosby:2013bix} 
%claims an upper limit $|a_{cn}|<2.8$ fm with a very small best fit value $a_{cn}=-0.05$ fm,
finds no low-lying $s$-wave $cn$ resonance, Ref.~\cite{Leblond:2015} claims 
%that the $cn$ subsystem has 
one with an energy $E_r\sim 0.8$ MeV and a width $\Gamma$ of a similar size.  One may ask if this resonance alone can lead to a bound $^{22}$C.  As both $E_r$ and $\Gamma$ are much larger than the energy of the $nn$ virtual state, we can use our previous results at unitarity.  For $A=20$ the $cnn$ bound state starts to appear at $r_{cn}/a_{cn}=1.929$, implying a resonance pole at $E-\frac i2 \Gamma$ with $\Gamma/E=3.64$.  
%This is larger the value measured in Ref.~\cite{Leblond:2015}.  
Thus, for the bound $^{22}$C to be explained solely by an $s$-wave $n+^{20}$C resonance, the latter has to be substantially wider than reported in Ref.~\cite{Leblond:2015}.

\emph{Summary.}---In this work we have demonstrated that, within an EFT framework, two‑body information alone suffices to describe two‑neutron halo nuclei, provided that the core–neutron ($cn$) scattering length $a_{cn}$ and effective range $r_{cn}$ are resummed to all orders. Using this nonperturbative treatment we calculated a set of observables that characterize the halo structure and track their systematic dependence on the core mass number $A$.

The formalism is most relevant when the $cn$ subsystem supports a low‑lying s-wave resonance. As a concrete application, we analyzed the Borromean nucleus $^{22}$C, extracting the requirements on the resonance position in the $cn$ channel so that it would yield a bound state in the $cnn$ system. We found that the resonance parameters given in Ref.~\cite{Leblond:2015} do not satisfy the requirements for our model to give a bound $^{22}$C.

Our power counting assumes that $r_{cn}$ is parametrically larger than the remaining effective‑range parameters, justifying its resummation. This leads to qualitatively different predictions from those of the standard short‑range EFT with perturbative range corrections~\cite{Ji:2011qg}. Distinguishing these two scenarios unambiguously will require either precise $cn$ data or multiple, well‑determined $cnn$ observables that can be confronted with the present EFT predictions.  Near the three-body threshold our model should reproduce the results of Ref.~\cite{Hongo:2022sdr} and provide an estimate for the correction due to $cn$ scattering.  We plan to investigate this in future work.

\begin{acknowledgments}
This work was supported by the National Science Foundation (Grant Nos.\ PHY-2111426 and PHY-2412612), the Office of Nuclear Physics, and the  US Department of Energy (Contracts Nos.\ DE-AC05-00OR22725 and DE-SC0009924).  D.T.S. thanks Takashi Nakamura for discussion and the Institut des Hautes \'Etudes Scientifiques, where part of this work was completed, for hospitality.
\end{acknowledgments}


\begin{thebibliography}{18}%
\makeatletter
\providecommand \@ifxundefined [1]{%
 \@ifx{#1\undefined}
}%
\providecommand \@ifnum [1]{%
 \ifnum #1\expandafter \@firstoftwo
 \else \expandafter \@secondoftwo
 \fi
}%
\providecommand \@ifx [1]{%
 \ifx #1\expandafter \@firstoftwo
 \else \expandafter \@secondoftwo
 \fi
}%
\providecommand \natexlab [1]{#1}%
\providecommand \enquote  [1]{``#1''}%
\providecommand \bibnamefont  [1]{#1}%
\providecommand \bibfnamefont [1]{#1}%
\providecommand \citenamefont [1]{#1}%
\providecommand \href@noop [0]{\@secondoftwo}%
\providecommand \href [0]{\begingroup \@sanitize@url \@href}%
\providecommand \@href[1]{\@@startlink{#1}\@@href}%
\providecommand \@@href[1]{\endgroup#1\@@endlink}%
\providecommand \@sanitize@url [0]{\catcode `\\12\catcode `\$12\catcode
  `\&12\catcode `\#12\catcode `\^12\catcode `\_12\catcode `\%12\relax}%
\providecommand \@@startlink[1]{}%
\providecommand \@@endlink[0]{}%
\providecommand \url  [0]{\begingroup\@sanitize@url \@url }%
\providecommand \@url [1]{\endgroup\@href {#1}{\urlprefix }}%
\providecommand \urlprefix  [0]{URL }%
\providecommand \Eprint [0]{\href }%
\providecommand \doibase [0]{https://doi.org/}%
\providecommand \selectlanguage [0]{\@gobble}%
\providecommand \bibinfo  [0]{\@secondoftwo}%
\providecommand \bibfield  [0]{\@secondoftwo}%
\providecommand \translation [1]{[#1]}%
\providecommand \BibitemOpen [0]{}%
\providecommand \bibitemStop [0]{}%
\providecommand \bibitemNoStop [0]{.\EOS\space}%
\providecommand \EOS [0]{\spacefactor3000\relax}%
\providecommand \BibitemShut  [1]{\csname bibitem#1\endcsname}%
\let\auto@bib@innerbib\@empty
%</preamble>
\bibitem [{\citenamefont {Tanihata}(1996)}]{Tanihata_1996}%
  \BibitemOpen
  \bibfield  {author} {\bibinfo {author} {\bibfnamefont {I.}~\bibnamefont
  {Tanihata}},\ }\bibfield  {title} {\bibinfo {title} {Neutron halo nuclei},\
  }\href {https://doi.org/10.1088/0954-3899/22/2/004} {\bibfield  {journal}
  {\bibinfo  {journal} {J. Phys. G: Nucl. Part. Phys.}\ }\textbf {\bibinfo
  {volume} {22}},\ \bibinfo {pages} {157} (\bibinfo {year} {1996})}\BibitemShut
  {NoStop}%
\bibitem [{\citenamefont {Braaten}\ and\ \citenamefont
  {Hammer}(2006)}]{Braaten:2004rn}%
  \BibitemOpen
  \bibfield  {author} {\bibinfo {author} {\bibfnamefont {E.}~\bibnamefont
  {Braaten}}\ and\ \bibinfo {author} {\bibfnamefont {H.~W.}\ \bibnamefont
  {Hammer}},\ }\bibfield  {title} {\bibinfo {title} {{Universality in few-body
  systems with large scattering length}},\ }\href
  {https://doi.org/10.1016/j.physrep.2006.03.001} {\bibfield  {journal}
  {\bibinfo  {journal} {Phys. Rep.}\ }\textbf {\bibinfo {volume} {428}},\
  \bibinfo {pages} {259} (\bibinfo {year} {2006})},\ \Eprint
  {https://arxiv.org/abs/cond-mat/0410417} {arXiv:cond-mat/0410417}
  \BibitemShut {NoStop}%
\bibitem [{\citenamefont {Efimov}(1991)}]{PhysRevC.44.2303}%
  \BibitemOpen
  \bibfield  {author} {\bibinfo {author} {\bibfnamefont {V.}~\bibnamefont
  {Efimov}},\ }\bibfield  {title} {\bibinfo {title} {Force-range correction in
  the three-body problem: Application to three-nucleon systems},\ }\href
  {https://doi.org/10.1103/PhysRevC.44.2303} {\bibfield  {journal} {\bibinfo
  {journal} {Phys. Rev. C}\ }\textbf {\bibinfo {volume} {44}},\ \bibinfo
  {pages} {2303} (\bibinfo {year} {1991})}\BibitemShut {NoStop}%
\bibitem [{\citenamefont {Bedaque}\ \emph {et~al.}(1999)\citenamefont
  {Bedaque}, \citenamefont {Hammer},\ and\ \citenamefont {van
  Kolck}}]{Bedaque:1998kg}%
  \BibitemOpen
  \bibfield  {author} {\bibinfo {author} {\bibfnamefont {P.~F.}\ \bibnamefont
  {Bedaque}}, \bibinfo {author} {\bibfnamefont {H.~W.}\ \bibnamefont
  {Hammer}},\ and\ \bibinfo {author} {\bibfnamefont {U.}~\bibnamefont {van
  Kolck}},\ }\bibfield  {title} {\bibinfo {title} {{Renormalization of the
  Three-Body System with Short-Range Interactions}},\ }\href
  {https://doi.org/10.1103/PhysRevLett.82.463} {\bibfield  {journal} {\bibinfo
  {journal} {Phys. Rev. Lett.}\ }\textbf {\bibinfo {volume} {82}},\ \bibinfo
  {pages} {463} (\bibinfo {year} {1999})},\ \Eprint
  {https://arxiv.org/abs/nucl-th/9809025} {arXiv:nucl-th/9809025} \BibitemShut
  {NoStop}%
\bibitem [{\citenamefont {Bedaque}\ \emph {et~al.}(2000)\citenamefont
  {Bedaque}, \citenamefont {Hammer},\ and\ \citenamefont {van
  Kolck}}]{Bedaque:1999ve}%
  \BibitemOpen
  \bibfield  {author} {\bibinfo {author} {\bibfnamefont {P.~F.}\ \bibnamefont
  {Bedaque}}, \bibinfo {author} {\bibfnamefont {H.~W.}\ \bibnamefont
  {Hammer}},\ and\ \bibinfo {author} {\bibfnamefont {U.}~\bibnamefont {van
  Kolck}},\ }\bibfield  {title} {\bibinfo {title} {{Effective theory of the
  triton}},\ }\href {https://doi.org/10.1016/S0375-9474(00)00205-0} {\bibfield
  {journal} {\bibinfo  {journal} {Nucl. Phys. A}\ }\textbf {\bibinfo {volume}
  {676}},\ \bibinfo {pages} {357} (\bibinfo {year} {2000})},\ \Eprint
  {https://arxiv.org/abs/nucl-th/9906032} {arXiv:nucl-th/9906032} \BibitemShut
  {NoStop}%
\bibitem [{\citenamefont {Hammer}\ \emph {et~al.}(2017)\citenamefont {Hammer},
  \citenamefont {Ji},\ and\ \citenamefont {Phillips}}]{Hammer:2017tjm}%
  \BibitemOpen
  \bibfield  {author} {\bibinfo {author} {\bibfnamefont {H.~W.}\ \bibnamefont
  {Hammer}}, \bibinfo {author} {\bibfnamefont {C.}~\bibnamefont {Ji}},\ and\
  \bibinfo {author} {\bibfnamefont {D.~R.}\ \bibnamefont {Phillips}},\
  }\bibfield  {title} {\bibinfo {title} {{Effective field theory description of
  halo nuclei}},\ }\href {https://doi.org/10.1088/1361-6471/aa83db} {\bibfield
  {journal} {\bibinfo  {journal} {J. Phys. G: Nucl. Part. Phys.}\ }\textbf
  {\bibinfo {volume} {44}},\ \bibinfo {pages} {103002} (\bibinfo {year}
  {2017})},\ \Eprint {https://arxiv.org/abs/1702.08605} {arXiv:1702.08605}
  \BibitemShut {NoStop}%
\bibitem [{\citenamefont {Mosby}\ \emph {et~al.}(2013)\citenamefont {Mosby}
  \emph {et~al.}}]{Mosby:2013bix}%
  \BibitemOpen
  \bibfield  {author} {\bibinfo {author} {\bibfnamefont {S.}~\bibnamefont
  {Mosby}} \emph {et~al.},\ }\bibfield  {title} {\bibinfo {title} {{Search for
  $^{21}$C and constraints on $^{22}$C}},\ }\href
  {https://doi.org/10.1016/j.nuclphysa.2013.04.004} {\bibfield  {journal}
  {\bibinfo  {journal} {Nucl. Phys.}\ }\textbf {\bibinfo {volume} {A909}},\
  \bibinfo {pages} {69} (\bibinfo {year} {2013})},\ \Eprint
  {https://arxiv.org/abs/1304.4507} {arXiv:1304.4507} \BibitemShut {NoStop}%
\bibitem [{\citenamefont {Hongo}\ and\ \citenamefont
  {Son}(2022)}]{Hongo:2022sdr}%
  \BibitemOpen
  \bibfield  {author} {\bibinfo {author} {\bibfnamefont {M.}~\bibnamefont
  {Hongo}}\ and\ \bibinfo {author} {\bibfnamefont {D.~T.}\ \bibnamefont
  {Son}},\ }\bibfield  {title} {\bibinfo {title} {{Universal Properties of
  Weakly Bound Two-Neutron Halo Nuclei}},\ }\href
  {https://doi.org/10.1103/PhysRevLett.128.212501} {\bibfield  {journal}
  {\bibinfo  {journal} {Phys. Rev. Lett.}\ }\textbf {\bibinfo {volume} {128}},\
  \bibinfo {pages} {212501} (\bibinfo {year} {2022})},\ \Eprint
  {https://arxiv.org/abs/2201.09912} {arXiv:2201.09912} \BibitemShut {NoStop}%
\bibitem [{\citenamefont {Petrov}(2004)}]{PhysRevLett.93.143201}%
  \BibitemOpen
  \bibfield  {author} {\bibinfo {author} {\bibfnamefont {D.~S.}\ \bibnamefont
  {Petrov}},\ }\bibfield  {title} {\bibinfo {title} {{Three-Boson Problem near
  a Narrow Fesh\-bach Resonance}},\ }\href
  {https://doi.org/10.1103/PhysRevLett.93.143201} {\bibfield  {journal}
  {\bibinfo  {journal} {Phys. Rev. Lett.}\ }\textbf {\bibinfo {volume} {93}},\
  \bibinfo {pages} {143201} (\bibinfo {year} {2004})},\ \Eprint
  {https://arxiv.org/abs/cond-mat/0404036} {arXiv:cond-mat/0404036}
  \BibitemShut {NoStop}%
\bibitem [{\citenamefont {Habashi}\ \emph {et~al.}(2021)\citenamefont
  {Habashi}, \citenamefont {Fleming},\ and\ \citenamefont {van
  Kolck}}]{Habashi:2020ofb}%
  \BibitemOpen
  \bibfield  {author} {\bibinfo {author} {\bibfnamefont {J.~B.}\ \bibnamefont
  {Habashi}}, \bibinfo {author} {\bibfnamefont {S.}~\bibnamefont {Fleming}},\
  and\ \bibinfo {author} {\bibfnamefont {U.}~\bibnamefont {van Kolck}},\
  }\bibfield  {title} {\bibinfo {title} {{Nonrelativistic effective field
  theory with a resonance field}},\ }\href
  {https://doi.org/10.1140/epja/s10050-021-00452-5} {\bibfield  {journal}
  {\bibinfo  {journal} {Eur. Phys. J. A}\ }\textbf {\bibinfo {volume} {57}},\
  \bibinfo {pages} {169} (\bibinfo {year} {2021})},\ \Eprint
  {https://arxiv.org/abs/2012.14995} {arXiv:2012.14995} \BibitemShut {NoStop}%
\bibitem [{\citenamefont {Habashi}\ \emph {et~al.}(2020)\citenamefont
  {Habashi}, \citenamefont {Sen}, \citenamefont {Fleming},\ and\ \citenamefont
  {van Kolck}}]{Habashi:2020qgw}%
  \BibitemOpen
  \bibfield  {author} {\bibinfo {author} {\bibfnamefont {J.~B.}\ \bibnamefont
  {Habashi}}, \bibinfo {author} {\bibfnamefont {S.}~\bibnamefont {Sen}},
  \bibinfo {author} {\bibfnamefont {S.}~\bibnamefont {Fleming}},\ and\ \bibinfo
  {author} {\bibfnamefont {U.}~\bibnamefont {van Kolck}},\ }\bibfield  {title}
  {\bibinfo {title} {{Effective field theory for two-body systems with shallow
  $S$-wave resonances}},\ }\href {https://doi.org/10.1016/j.aop.2020.168283}
  {\bibfield  {journal} {\bibinfo  {journal} {Ann. Phys. (NY)}\ }\textbf
  {\bibinfo {volume} {422}},\ \bibinfo {pages} {168283} (\bibinfo {year}
  {2020})},\ \Eprint {https://arxiv.org/abs/2007.07360} {arXiv:2007.07360}
  \BibitemShut {NoStop}%
\bibitem [{\citenamefont {Nishida}(2012)}]{Nishida:2012xw}%
  \BibitemOpen
  \bibfield  {author} {\bibinfo {author} {\bibfnamefont {Y.}~\bibnamefont
  {Nishida}},\ }\bibfield  {title} {\bibinfo {title} {{New Type of Crossover
  Physics in Three-Component Fermi Gases}},\ }\href
  {https://doi.org/10.1103/PhysRevLett.109.240401} {\bibfield  {journal}
  {\bibinfo  {journal} {Phys. Rev. Lett.}\ }\textbf {\bibinfo {volume} {109}},\
  \bibinfo {pages} {240401} (\bibinfo {year} {2012})},\ \Eprint
  {https://arxiv.org/abs/1207.6971} {arXiv:1207.6971} \BibitemShut {NoStop}%
\bibitem [{\citenamefont {Griesshammer}\ and\ \citenamefont {van
  Kolck}(2023)}]{Griesshammer:2023scn}%
  \BibitemOpen
  \bibfield  {author} {\bibinfo {author} {\bibfnamefont {H.~W.}\ \bibnamefont
  {Griesshammer}}\ and\ \bibinfo {author} {\bibfnamefont {U.}~\bibnamefont {van
  Kolck}},\ }\bibfield  {title} {\bibinfo {title} {{Universality of three
  identical bosons with large, negative effective range}},\ }\href
  {https://doi.org/10.1140/epja/s10050-023-01196-0} {\bibfield  {journal}
  {\bibinfo  {journal} {Eur. Phys. J. A}\ }\textbf {\bibinfo {volume} {59}},\
  \bibinfo {pages} {289} (\bibinfo {year} {2023})},\ \Eprint
  {https://arxiv.org/abs/2308.01394} {arXiv:2308.01394} \BibitemShut {NoStop}%
\bibitem [{\citenamefont {Vanasse}(2017)}]{Vanasse:2016hgn}%
  \BibitemOpen
  \bibfield  {author} {\bibinfo {author} {\bibfnamefont {J.}~\bibnamefont
  {Vanasse}},\ }\bibfield  {title} {\bibinfo {title} {{Charge and matter form
  factors of two-neutron halo nuclei in halo effective field theory at
  next-to-leading order}},\ }\href {https://doi.org/10.1103/PhysRevC.95.024318}
  {\bibfield  {journal} {\bibinfo  {journal} {Phys. Rev. C}\ }\textbf {\bibinfo
  {volume} {95}},\ \bibinfo {pages} {024318} (\bibinfo {year} {2017})},\
  \Eprint {https://arxiv.org/abs/1609.08552} {arXiv:1609.08552} \BibitemShut
  {NoStop}%
\bibitem [{slo()}]{slope}%
  \BibitemOpen
  \href@noop {} {}\bibinfo {note} {The slopes $\alpha(A)$ are the following:
  $\alpha(1) = 0.81$, $\alpha(2) = 1.01$, $\alpha(3)= 1.19$, $\alpha(5) =
  1.43$, $\alpha(8) = 1.63$, $\alpha(10) = 1.72$, $\alpha(20) = 1.93$,
  $\alpha(30) = 2.01$, $\alpha(\infty) = 2.20$.}\BibitemShut {Stop}%
\bibitem [{\citenamefont {Konishi}\ \emph {et~al.}(2018)\citenamefont
  {Konishi}, \citenamefont {Morimatsu},\ and\ \citenamefont
  {Yasui}}]{Konishi:2017lbg}%
  \BibitemOpen
  \bibfield  {author} {\bibinfo {author} {\bibfnamefont {A.}~\bibnamefont
  {Konishi}}, \bibinfo {author} {\bibfnamefont {O.}~\bibnamefont {Morimatsu}},\
  and\ \bibinfo {author} {\bibfnamefont {S.}~\bibnamefont {Yasui}},\ }\bibfield
   {title} {\bibinfo {title} {{Degenerate two-body and three-body
  coupled-channels systems: Renormalized effective Alt-Grassberger-Sandhas
  equations and near-threshold resonances}},\ }\href
  {https://doi.org/10.1103/PhysRevC.97.064001} {\bibfield  {journal} {\bibinfo
  {journal} {Phys. Rev. C}\ }\textbf {\bibinfo {volume} {97}},\ \bibinfo
  {pages} {064001} (\bibinfo {year} {2018})},\ \Eprint
  {https://arxiv.org/abs/1703.04073} {arXiv:1703.04073} \BibitemShut {NoStop}%
\bibitem [{\citenamefont {Leblond}(2015)}]{Leblond:2015}%
  \BibitemOpen
  \bibfield  {author} {\bibinfo {author} {\bibfnamefont {S.}~\bibnamefont
  {Leblond}},\ }\emph {\bibinfo {title} {{Structure des isotopes de bore et de
  carbone riches en neutrons aux limites de la stabilit\'e}}},\ \href
  {https://theses.hal.science/tel-01289381} {Ph.D. thesis},\ \bibinfo  {school}
  {University of Caen Normandie} (\bibinfo {year} {2015})\BibitemShut {NoStop}%
\bibitem [{\citenamefont {Ji}\ \emph {et~al.}(2012)\citenamefont {Ji},
  \citenamefont {Phillips},\ and\ \citenamefont {Platter}}]{Ji:2011qg}%
  \BibitemOpen
  \bibfield  {author} {\bibinfo {author} {\bibfnamefont {C.}~\bibnamefont
  {Ji}}, \bibinfo {author} {\bibfnamefont {D.~R.}\ \bibnamefont {Phillips}},\
  and\ \bibinfo {author} {\bibfnamefont {L.}~\bibnamefont {Platter}},\
  }\bibfield  {title} {\bibinfo {title} {{The three-boson system at
  next-to-leading order in an effective field theory for systems with a large
  scattering length}},\ }\href {https://doi.org/10.1016/j.aop.2012.02.001}
  {\bibfield  {journal} {\bibinfo  {journal} {Ann. Phys. (NY)}\ }\textbf
  {\bibinfo {volume} {327}},\ \bibinfo {pages} {1803} (\bibinfo {year}
  {2012})},\ \Eprint {https://arxiv.org/abs/1106.3837} {arXiv:1106.3837}
  \BibitemShut {NoStop}%
\end{thebibliography}
\end{document}